\documentclass[reprint,aps,prl,notitlepage]{revtex4-2}

\usepackage{graphicx}
\usepackage{overpic}

\begin{document}

\pagestyle{empty}

\textbf{Comment on ``Anomalous Reentrant 5/2 Quantum Hall Phase at Moderate Landau-Level-Mixing Strength''} 

\vspace*{5pt}

Das, Das, and Mandal\cite{Das} examine a wavefunction for $\nu=5/2$ on a sphere including moderate Landau-Level mixing evaluated perturbatively.  In contrast to the claims of Ref.~\cite{Das}, the wavefunction they find is not a fractional quantum Hall (FQH) state.     Splitting the system into two hemispheres each with $N/2$ particles, Ref.~\cite{Das} shows entanglement spectra (Figs.~3, 5) which    (i) do not change qualitatively when two flux are either added or subtracted and (ii) have a branch that remains at very low entanglement energy (high weight) out to the maximum possible $L_z$ angular momentum.  Neither (i) nor (ii) is the case for any known FQH states.   Instead, I claim that these wavefunctions  exhibit  phase separation or bubble/stripe formation.   

Assuming the system of $N$ electrons separates into a compact filled region ($\nu=1$) and a remaining empty region ($\nu=0$), on a sphere we would expect a ground state at maximal angular momentum.  If we focus instead on  angular momentum $L^2=0$ states, the best clustering occurs by separating the system into two filled regions containing $N/2$ electrons each (assuming even $N$),  which are arranged antipodally (opposite) on the sphere, and then we sum over all directions of the antipodal axis to obtain $L^2=0$.   I claim that this $L^2=0$ antipodal cluster state is what is found in the numerics of Ref.~\cite{Das}.    

\begin{figure}[h!]

\vspace*{1pt}

\includegraphics[width=\columnwidth]{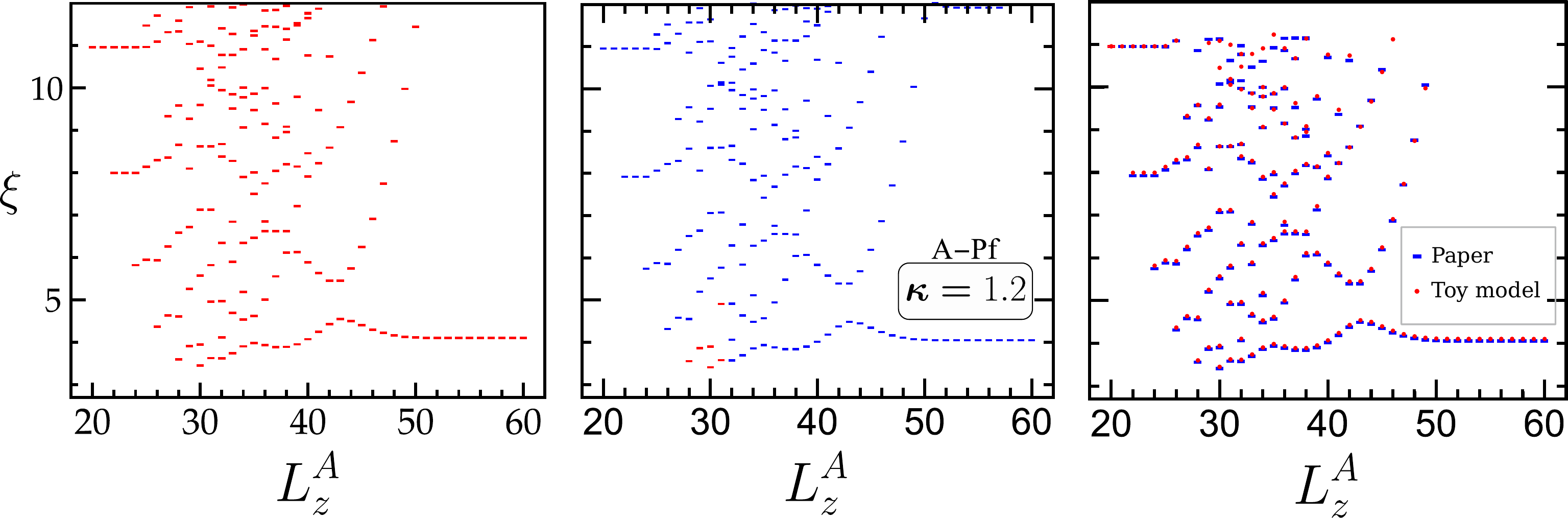}
\caption{Entanglement spectra for two $N=12$ electron states on a sphere with $N_\phi=25$ flux.  The entanglement cut is along the equator and each half of the cut has $N/2$ electrons.  ({\it Left}) Toy model.  The wavefunction  is the lowest energy $L^2=0$ state of a Hamiltonian with short range attraction: $V_1 = -1$ and $V_{n\neq 1} = 0$.  ({\it Middle}) Data taken from Fig.~3c of Ref.~\cite{Das}. The wavefunction is the ground state with Landau-level mixing parameter $\kappa =1.2$ using the approximate Hamiltonian described in Ref.~\cite{Das}.  ({\it Right})  left and middle overlayed. }
\end{figure}

This type of antipodal clustering should not care much about the precise value of $N$ (even) or flux $N_\phi$ in agreement with (i).   Further, such a wavefunction should have high entanglement for a cut with maximal $L_z$ from the configuration with clusters at the north and south poles.   This is 
in agreement with (ii).  A branch of lower $L_z$ states should also have similarly high entanglement,  corresponding to configurations with the antipodal axis not along the poles.  These statements are in agreement with the numerical observations of Ref.~\cite{Das}. 

In order to prove our claim that the data of Ref.~\cite{Das} is showing clustering, following Ref.~\cite{Mila} we consider electrons with short ranged attraction only: i.e., Haldane pseudopotentials $V_1 = -1$ and $V_{n \neq 1} =0$.   With such a Hamiltonian the electrons attract each other as much as possible, forming entirely filled regions.  The ground state is at maximum $L$, but if we look instead at the lowest energy $L^2=0$ state, we see in Fig.~1 that its entanglement spectrum is identical to that shown in Ref.~\cite{Das}.  Note that this particular toy-model Hamiltonian is particle-hole symmetric, and for filling $\nu < 1/2$ the holes will cluster instead of the electrons.  The Hamiltonian considered in Ref.~\cite{Das}, in contrast, is not particle-hole symmetric and shows the same electron clustering behavior for all fillings shown in Fig. 3.  

The explicit analytic wavefunction shown in Ref.~\cite{Das} (Eq. 2) is a Halperin 113 state\cite{Halperin} which has been fully antisymmetrized between two species.   It is known that the 113 state phase separates between species\cite{deGail}.  Once the species are physically separated, the effects of antisymmetrization are expected to be minor.   Here, the phase separation is so strong that the electrons form regions that are entirely filled and entirely empty. 
 
The fact that the system breaks up into antipodal clusters is perhaps not surprising given that there is a well-known tendency to stripe/bubble formation in the higher Landau levels\cite{Fogler} and at $\nu=5/2$ when strong enough Landau level mixing is included perturbatively\cite{Mila}.   What might be more surprising is that in all cases observed in Ref.~\cite{Das}, it appears there are exactly two clusters on the sphere. There could be several reasons for this.  First, if it were just a single cluster, one would not have $L^2=0$.
Secondly, it is possible that such finite size systems are too small to fit more than two clusters.

In summary, the states studied in Ref.~\cite{Das} are not FQH states but rather show phase separations or stripe/bubble formation.  All conclusions stated by Ref.~\cite{Das} must be considered in this light. 

I thank G. J. Sreejith for providing numerical data in Fig. 1.   Funding from EPSRC EP/S020527/1.

\vspace*{10pt}

\noindent Steven H. Simon

\vspace*{5pt}

\hspace*{10pt}   \begin{minipage}{3in}
    \small Rudolf Peierls Center for Theoretical Physics\\
    Oxford, United Kingdom, OX1 3PU
   \end{minipage}

\end{document}